%% file: GravitinoDM.tex
\documentclass[aps,prd,floatfix,nofootinbib,preprintnumbers]{revtex4}
\usepackage[english]{babel}
\usepackage{amsmath,amssymb}
\usepackage{graphicx}
\usepackage{graphics}

\def\lsim{\raise0.3ex\hbox{$\;<$\kern-0.75em\raise-1.1ex\hbox{$\sim\;$}}}

\newcommand{\eq}[1]{eq.~(\ref{eq:#1})}
\newcommand{\fig}[1]{Fig.~\ref{fig:#1}}
\newcommand{\AddrWur}{%
Institut f\"ur Theoretische Physik und Astronomie, 
Universit\"at W\"urzburg\\
Am Hubland, 
97074 W\"urzburg}
\newcommand{\AddrGot}{%
Institut f\"ur Astrophysik, 
Universit\"at G\"ottingen \\
Friedrich-Hund-Platz 1, 
37077 G\"ottingen}

\begin{document}

\title{Strong dark matter constraints on GMSB models}

\author{F.\ Staub} \email{florian.staub@physik.uni-wuerzburg.de}
\affiliation{\AddrWur}

\author{J.\ Niemeyer} \email{niemeyer@astro.physik.uni-goettingen.de} \affiliation{\AddrGot}

\author{W.\ Porod} \email{porod@physik.uni-wuerzburg.de} \affiliation{\AddrWur}


\begin{abstract}
We reconsider the dark matter problem in supersymmetric
models with gauge mediated supersymmetry breaking, with and without $R$-parity breaking. In
these classes of models, a light gravitino forms the dark matter.
Consistency with the experimental data, in particular  
the dark matter abundance and the small-scale power spectrum,
requires additional entropy
production after the decoupling of the gravitino from the thermal bath.  We demonstrate that 
the usual mechanism via messenger number violating interactions does not work in models where
the messenger belongs to $SU(5)$ representations. This is mainly a
consequence of two facts: (i) there are at least two different types of lightest
messenger particles and (ii) the lightest messenger particle with $SU(2)$  
quantum numbers decays
dominantly into vector bosons once messenger number is broken, a feature which has been overlooked so
far. 
In case of $SO(10)$ messenger multiplets we find scenarios which work if the SM gauge singlet
component is rather light.
\end{abstract}

\maketitle

\section{Introduction}

One of the most attractive features of the minimal supersymmetric
standard model (MSSM) is the existence of a stable particle, usually
the lightest neutralino, which successfully serves as a dark matter
candidate yielding the correct amount of dark matter required by
observations. Models in which the effects of supersymmetry (SUSY)
breaking are communicated via the usual gauge interactions
\cite{Dine:1981gu,Dine:1981za,Dimopoulos:1981au,Nappi:1982hm,AlvarezGaume:1981wy,Dine:1993yw,Dine:1993qm,Dine:1994vc,Dine:1995ag} to the
''visible'' have recently achieved considerable attention
(see e.g.~\cite{Intriligator:2006dd,Murayama:2006yf,Murayama:2007fe,Carpenter:2008wi,%
Dienes:2008gj,Komargodski:2008ax} and references therein), in particular in view
of model building and understanding the mechanism of supersymmetry breaking. 
An attractive feature of such models is the natural
explanation for the smallness of SUSY contributions to
flavour-changing neutral current phenomena due to the strongly
constrained SUSY spectrum.

Before discussing this model in more detail, let us briefly recall
important early universe issues which are usually more or less
implicitly assumed when discussing dark matter
constraints within the MSSM framework:
 (i) The particle content of the universe at the end of inflation is
assumed to be described by the MSSM plus graviton and gravitino. This
implies that the hidden sector responsible for supersymmetry breaking
and/or its communication to the MSSM is essentially heavier than the
reheating temperature. As a consequence, it is not produced during the
reheating phase and has no impact on the later evolution of the
universe.
(ii) All MSSM particles are initially in thermal equilibrium.
(iii) The gravitino may or may not be in thermal equilibrium. In the
former case its number density depends strongly on the reheating
temperature.
Items (i) and (ii) allow, for example, a routine relic density
calculation for thermally produced dark matter in minimal supergravity (mSUGRA) scenarios with a
neutralino as the lightest supersymmetric particle (LSP)
 (and also in similar scenarios).  Item (iii) often
creates problems: in scenarios where the gravitino is the 
 LSP, the next to lightest supersymmetric
particle (NLSP) can be rather long-lived with life times up to several
years. In scenarios where the gravitino is not the LSP, it itself can be
similarly long-lived. In both cases one has to check that the
corresponding decays do not spoil the successful predictions of
big-bang nucleosynthesis (BBN).

In gauge mediated SUSY breaking (GMSB) models there are two new
sectors \cite{Dine:1993yw,Dine:1993qm,Dine:1994vc,Dine:1995ag,Giudice:1998bp}:
\begin{enumerate}
\item the secluded sector: This is a strongly interacting
sector in which SUSY is broken dynamically. 
\item the messenger sector: It contains fields charged under
  $SU(3)\times SU(2) \times U(1)$ gauge interactions which communicate
  SUSY breaking to the ordinary sparticles. Usually one assumes that
  they come in complete $SU(5)$ representations or representations of
  larger groups containing $SU(5)$ as subgroup, so that the success of
  gauge coupling unification does not get spoiled.
\end{enumerate}
In these scenarios the gravitino is usually very light, in the range
between a few eV up to O(1) MeV. The masses of the messenger particles
and some of the fields in the secluded sector can be as low as 100 TeV
implying that they can act as cold dark matter if their masses are below
the reheat temperature and if they are stable.

The gravitino is the lightest SUSY particle in GMSB models and all
MSSM particles decay into it in a cosmologically short time. Therefore,
the gravitino forms the dark matter in these models. This aspect has
been extensively discussed in the literature
\cite{Dimopoulos:1996gy,Han:1997wn,Baltz:2001rq,Fujii:2002fv,Fujii:2002yx,Jedamzik:2005ir}.
The abundance of thermal produced gravitinos is under assumptions
consistent with the standard thermal evolution of the early universe given by 
\begin{equation}
\label{eq:relic}
\Omega_{3/2} h^2 = \frac{m_{3/2}}{\mbox{keV}} \frac{100}{g_\star} \, \, .
\end{equation} 
Here \(g_\star\) is the effective number of degrees of freedom at the time of gravitino decoupling. 
This implies that the gravitino forms warm dark matter (WDM) in the
mass range of $O(100)$ eV.

However, there are stringent
constraints on the contribution of WDM particles 
with free-streaming lengths of the order of galaxy scales or larger to
the total dark matter content. More precisely, if dark matter is
assumed to consist of only one particle species, its mass is limited
from below by the amplitude of the small-scale power spectrum which,
in turn, currently receives its tightest constraints from observations of the
Lyman-\(\alpha\) forest \cite{Viel:2005qj}. This bound has recently
been increased by an order of magnitude ruling out pure WDM scenarios
with particle masses below 8 keV for non resonantly produced dark matter \cite{Boyarsky:2008xj}. 
Gravitinos with masses up to \(O(\mbox{MeV})\) have once been in thermal equilibrium 
as long as the reheating temperature is above \(10^6 \mbox{GeV}\).   
For those thermal relics the bound is 1.5 keV, while mixed dark
matter scenarios dominated by a cold component allow a contribution of
up to 60\% by a WDM particle of any mass above 1.1 keV. 
There remain a number of 
systematic uncertainties in the interpretation of Lyman-\(\alpha\)
observations, most of which related to the
poorly understood thermal evolution of the intergalactic medium, but
the overall result is fairly robust (for a detailed discussion, see
\cite{Boyarsky:2008xj}), so that pure gravitino dark matter allowed by Lyman-\(\alpha\) bounds would have a relic density at least 15 times higher than the measured dark matter relic density. \\
An additional difficulty stems from the fact that in typical GMSB
models, the lightest messenger particle is stable as a result of the
conservation of a messenger quantum number. Its relic density is
calculable similarly to the case of a neutralino LSP and is found to
scale as $\Omega_M h^2 \simeq 10^5 m_-^2/(10^3$~TeV$)^2$, where $m_-$
is the mass of the lightest messenger particle.  This overcloses the
universe in most of the parameter space as discussed below. The situation
becomes even worse because, as we will show below, there are usually several
different types of stable messenger particles, one for each type of
the corresponding SM gauge group representations.

A possible solution to both problems is additional entropy production
by decays of the lightest messenger particles into standard model
(SM) fermions
\cite{Han:1997wn,Baltz:2001rq,Fujii:2002fv,Fujii:2002yx,Jedamzik:2005ir}. The
basic idea is that, in general, gravitational interactions break global
symmetries and thus one expects terms like $f m_{3/2} \hat{\Phi}_M \hat{\bar{5}}$
in the superpotential \cite{Jedamzik:2005ir} where $\hat{\Phi}_M$ is a
messenger 5-plet and $\hat{\bar{5}}$ is a 5-plet containing the right down
quark superfield and the left lepton superfields of the MSSM. Owing to such
interactions, the lightest messenger field decays into standard model
fermions. It has been claimed in the literature that it is sufficiently long lived
to substantially produce entropy, diluting the gravitino
abundance. This in turn would imply that heavier gravitinos with
masses above 8 keV  would be viable DM candidates. 
However, we will show below that this statement is
incorrect  as only part of the
possible messenger decay modes have been taken into account in the literature.

So far we have implicitly assumed that $R$-parity is conserved implying
a stable LSP. There is one
experimental observation which can be explained by the breaking of
$R$-parity, namely neutrino masses and mixings. In the simplest model, 
 one adds bilinear $R$--parity breaking terms to the MSSM
superpotential and in this way neutrino data can be explained, see e.g.\
\cite{Romao:1999up,Hirsch:2000ef} and references therein. Moreover,
this class of models is also consistent with constraints from
baryogenesis \cite{Akeroyd:2003pb}.  Now, what about dark matter?
Neutrino physics gives a lower bound on the $R$-parity breaking
parameters such that the lightest MSSM particle will decay within a
small fraction of a second. However, this does not apply to a light
gravitino, which eventually decays, but neutrino physics now implies
that its life time is much larger than the age of universe
\cite{Borgani:1996ag,Takayama:2000uz,Hirsch:2005ag}.

The plan of this paper is as follows: in the next section we will
briefly discuss the main features of the models under
consideration. In section \ref{sec:decay} we will discuss the decays
of the lightest messenger pointing out the importance of decays into
vector bosons, which have not been discussed so far in the
literature. In section \ref{sec:dm} we discuss the consequences for
dark matter and in section \ref{sec:conc} we conclude.

\section{Models}
\label{sec:model}

Here, we briefly recall the main features of GMSB models needed for the
subsequent discussions.  Further details can be found in
ref.~\cite{Giudice:1998bp}. For the following discussion a minimal model is
sufficient, e.g., we assume one pair of 5-plets of messenger fields
coupled to one spurion superfield $\hat S$ in the secluded sector. The
resulting superpotential can be written as
\begin{eqnarray}
W = W_{MSSM} + \hat S \hat{\Phi}_M  \hat{\bar{\Phi}}_M + W_s(\hat S,\hat Z_i)
\label{eq:superpotential}
\end{eqnarray}
where $W_{MSSM}$ contains the MSSM superpotential and $W_s$ is the
superpotential of the secluded sector containing the fields $\hat S$
and $\hat Z_i$.  The field $\hat S$ is one of the fields responsible
for SUSY breaking and both the scalar component and the auxiliary
component receive vacuum expectation values (vevs) denoted by $M$ and
$F$, respectively. In the resulting effective action, $\hat S$ can be
replaced by $M + \theta\theta F$. As a consequence of SUSY breaking,
the messenger fermion receives a mass $M$ and the scalar components
obtain masses $m_{-,+} = M \sqrt{1 \mp F/M^2}$ and the corresponding
eigenstates are $\phi_- = (- \Phi^*_M + \bar{\Phi}_M )/\sqrt{2}$ and
$\phi_+ = (\Phi^*_M + \bar{\Phi}_M )/\sqrt{2}$.  Strictly speaking,
these formulas are valid at the GUT scale and different components get
renormalized differently due to differences in the corresponding gauge
couplings, resulting in a somewhat lighter $SU(2)$ doublet with the
same quantum numbers as the lepton doublet and a heavier $SU(3)$
triplet state carrying the same quantum numbers as the right-handed
$d$-quarks. However, the mass formula applies for both messenger types.
 Note that the requirement  $_->0$ implies $F \le
M^2$. Moreover, both the lightest d-type messenger scalar and the
lightest doublet scalar are stable due to  gauge invariance, e.g., there
are no kinematically allowed decay processes from one of these two fields into
the second one.

Since the messenger fields share the SM gauge
interactions, the gaugino and scalar SUSY particles obtain masses at
the one- and two-loop level \cite{Giudice:1998bp}, respectively:
\begin{equation}
M_i \sim \frac{\alpha_i}{4 \pi} \frac{F}{M} \,,\hspace{5mm}
m^2_x \sim \sum_a C_a \left(  \frac{\alpha_a}{4 \pi} \right)^2 \left(\frac{F}{M} \right)^2 \,\,,
\end{equation}
where the sum runs over the gauge couplings corresponding the quantum
numbers of the MSSM field $x$, and $C_a$ is the Casimir of the
corresponding representation. In order to obtain masses of the order
100 GeV to 1 TeV, one finds that $\Lambda \equiv F/M \sim 100$~TeV.
The gravitino mass is related to the fundamental SUSY breaking scale
\begin{equation}
m_{3/2} = \frac{F_{tot}}{\sqrt{3} m_{Pl}}  \,\,,
\label{eq:m32}
\end{equation}
where $F_{tot} = F + \sum_i F_{Z_i}$ is the sum of all $F$-terms in
the secluded sector and $m_{Pl}$ is the reduced Planck mass.  Defining
$k= F/F_{tot}$, one can write $m_{3/2} = F /(\sqrt{3} k m_{Pl})$.

Messenger gauge interactions and those derived from
\eq{superpotential} conserve messenger number so that the lightest
messenger boson is stable in this minimal version of the model.  In
scenarios where its mass is below the reheating temperature, this leads in
general to an over-closure of the universe
\cite{Dimopoulos:1996gy}. However, on general grounds one expects that
gravitational interactions either break this discrete symmetry
and/or induce higher-dimensional operators
\cite{Han:1997wn,Baltz:2001rq,Fujii:2002fv,Fujii:2002yx} implying that
the lightest messenger particle becomes unstable. A thorough
discussion of the various possibilities and the related theoretical
uncertainties can be found in ref.~\cite{Jedamzik:2005ir}. The general
conclusion has been that  in a large part of the parameter space,
the lightest messenger decays after the freeze out of the gravitino
and before BBN, thus allowing for a gravitino with masses of several
keV.  To be specific we consider the case 
\begin{equation}
\delta K = f m_{3/2} \hat{\Phi}_M \hat{\bar{5}}
\label{eq:deltaK}
\end{equation}
is added to the minimal K\"ahler potential $K_0 = \sum_i \hat{\Phi}^\dagger_i
\hat{\Phi}_i$ where the sum runs over all superfields. The constant $f$ is
usually assumed to be $O(1)$ but we will take it as a free parameter
allowing arbitrary values.  Using the usual invariance of the
supergravity Lagrangian under K\"ahler transformations $K\to K +
F(\hat{\Phi}) + F^*(\hat{\Phi}^*)$, $F=-\delta K$, followed by the superpotential
scaling $W \to e^{-F} W$ \cite{Wess:1992cp} one obtains in lowest order the following
additional contribution to the superpotential
\begin{equation}
\delta W = f m_{3/2} \hat{\Phi}_M \hat{\bar{5}}
\label{eq:deltaW}
\end{equation}
As we discuss in some detail later, this term leads to various decays
of the lightest messenger scalar.

The model(s) described so far conserve $R$-parity. As mentioned above,
one can explain neutrino data by breaking $R$-parity via lepton number
breaking terms. The simplest model consistent with the experimental
data is obtained by adding bilinear terms to the MSSM superpotential
\cite{Romao:1999up,Hirsch:2000ef}:
\begin{equation}
W_{MSSM} \to W_{MSSM} + \epsilon_i \hat L_i \hat H_u
\label{eq:bilinear}
\end{equation}
This way, one cannot only explain neutrino data but  also
predict certain decay properties of the LSP within the MSSM spectrum
in terms of neutrino mixing angles \cite{Porod:2000hv,Hirsch:2002ys,Hirsch:2003fe}. 
In our case, the
lightest SUSY particle is the gravitino and these predictions hold for
the NLSP \cite{Hirsch:2005ag}. Eventually also the gravitino will
decay but its life time is of the order of $10^{30}$ Hubble times
\cite{Borgani:1996ag,Takayama:2000uz,Hirsch:2005ag} and, thus, the
gravitino is in principle a valid DM candidate in this class of
models. An interesting question, which will be addressed below, is
whether the NLSP can be long lived enough so that its decay products
yield additional entropy production once the gravitino has been
decoupled from the thermal bath.

\section{Decay properties of messenger particles and the NLSP}
\label{sec:decay}

In GMSB models there are serious problems with cosmology as discussed
above: (i) if only the gravitino were responsible for dark matter
assuming the standard history of the universe, it would be warm dark
matter with a mass of about 100 eV, which is in conflict with the
measurements of the Lyman-$\alpha$ forest. (ii) If messenger particles
are produced after inflation and if messenger number is conserved, one
will in general obtain an exceedingly large contribution to $\Omega h^2$ which
overcloses the universe. Both problems can, in principle, be solved by
breaking messenger number, implying that the lightest messenger
particles decay and give rise to sufficient additional entropy
production once the gravitino is decoupled from the thermal bath,
which gives a lower bound on the life time of these particles. 

Let us first consider the spectrum of the models before discussing the
decay properties of the various particles. In the model with conserved
messenger number and conserved $R$-parity one has $m_+ > M > m_- \gg
m_{SUSY} \gg m_{3/2}$. Both, $f m_{3/2}$ in \eq{deltaW} and
$\epsilon_i$ in \eq{bilinear}, are small compared to $m_{SUSY}$
implying that the induced mixing will only give small corrections to
the various masses. This spectrum gives rise to decays of the
following type: $\phi_+ \to \tilde B \tilde \phi$ and $\tilde \phi \to
\tilde B \phi_-$, where $\tilde \phi$ is the messenger fermion. These
decays are so fast that $\phi_+$ and $\tilde \phi$ will decay
immediately after their decoupling from the thermal bath.

The term in \eq{deltaW} induces mixings between the messenger scalars
and the MSSM sfermions, in particular between messenger $SU(3)$
triplet and the squarks and between the messenger $SU(2)$ doublet and
the sleptons. For example, in the model with conserved $R$-parity one
obtains for the mixing matrix between the sleptons with $SU(2)$
doublet the following mass matrix in the basis $(\tilde L_i, \phi_-)$ where
$\tilde L$ are the left-sleptons:
\begin{equation}
\left( \begin{array}{cccc}
M^2_{L,11} + D + f^2 m^2_{3/2}& 0 & 0 & \frac{1}{\sqrt{2}} f m_{3/2} M \\
0 & M^2_{L,22} + D + f^2 m^2_{3/2}& 0 & \frac{1}{\sqrt{2}} f m_{3/2} M \\
0 & 0 & M^2_{L,33} + D + f^2 m^2_{3/2}& \frac{1}{\sqrt{2}} f m_{3/2} M \\
\frac{1}{\sqrt{2}} f m_{3/2} M  & \frac{1}{\sqrt{2}} f m_{3/2} M  & \frac{1}{\sqrt{2}} f m_{3/2} M 
& M^2 - F + D' + \frac{3}{2}f^2 m^2_{3/2} \\
\end{array}
\right)
\end{equation}
where $D$ and $D'$ denote  $D$-terms occurring after electro-weak symmetry breaking with
$D = O(m^2_Z)$ and $D'\simeq 0$. The latter is due to the fact $D'$ is propotional to
$\cos 2\theta$ where $\theta$ is the mixing angle between $\phi_-$ and $\phi_+$ which turns
out to be maximal: $\theta \simeq 1/\sqrt{2}$.  
For simplicity, we have assumed that the couplings of the messenger superfield
is generation independent, e.g.~$f_1=f_2=f_3=f$. Relaxing this assumption does not change
any of our conclusions.
The induced mixing between the sneutrinos and the neutral messenger scalar
is of the order
\begin{equation}
\delta \simeq \frac{f m_{3/2} M}{\sqrt{2} (M^2 - F)}
       \simeq \frac{f m_{3/2} M}{\sqrt{2} m^2_-}
\end{equation}

In the literature \cite{Baltz:2001rq,Fujii:2002fv,Jedamzik:2005ir} it is assumed that 
\(F \ll M^2\) and the
lightest messenger decays predominantly into a neutralino and a SM
fermion. The generic decay width is calculated to
\begin{equation}
\Gamma \simeq \frac{g^2}{16 \pi} \delta^2 m_-.
\label{eq:gamma_neut}
\end{equation}
This would imply that the lightest messenger scalars decay
after
the freeze out of the gravitinos and before BBN. Furthermore, the
dilution for a natural choice of parameters, e.g.~$f = O(1)$ could lead to a relic
density of the gravitinos of \(\Omega h^2 \simeq 0.1\).

This conclusion would be correct in the absence of an event which has not been
taken into account so far: electroweak symmetry breaking which opens new decay channels,
namely into vector-bosons. It has been shown  \cite{Bartl:1998xk,Bartl:1999bg} that
decays into $W$- and $Z$-bosons can dominate the decays of supersymmetric scalar particles
if there is sufficient phase space.
It turns out that in our scenario, the decay into $Z$ bosons are suppressed via
an extended  GIM mechanism. However, this is not the case for decays into $W$-bosons
and one gets for the  corresponding  widths 
\begin{equation}
\Gamma \simeq \frac{g^2}{16 \pi} \frac{m_-^2}{m_W^2}\delta^2 m_- .
\end{equation}  
The additional factor \(\frac{m_-^2}{m_W^2} \) implies that this
decay mode dominates once the lightest messenger scalar has a mass of a few TeV
and its life time gets reduced by the inverse factor. Therefore, it is of vital
importance to know how small $m_-$ can be as this gives an upper bound on the 
life-time. At tree level one might argue that \(\sqrt{M^2 - F} \simeq m_W\)
but it  turns out that the one-loop corrections dominate the mass in this case:
\begin{equation}
\delta m^2 = \sum_i \frac{1}{4 \pi^2} C^2_i g_i^2 M^2 .
\end{equation}
The sum is taken over contributing gauge couplings and \(C^2_i\) is
the corresponding Dynkin index. This can be easily understood by noting that
at the scale $M$, the messenger fermions decouple from the spectrum and, thus,
supersymmetry is broken as the number of bosonic degrees of freedom does not match
the number of fermionic degrees of freedom leading to this large correction.
Additionally the masses are increased
by the RGE running from the SUSY breaking scale to the low scale by a
factor \(\left(1-\frac{C_i^2 g_i^2(M_Z)}{8
  \pi^2}\ln\left(\frac{M^2}{M^2_Z}\right) \right)^{-1}\).

 We made a full analysis of the possible decays of the lightest
 messenger using the one loop corrections for the messenger
 masses. The SUSY parameters were calculated with SPheno
 \cite{Porod:2003um} and the \(\mu/B_\mu\) parameters were fixed by
 solving the tadpole equations.  All vertices and mass matrices of the
 model were derived with SARAH \cite{Staub:2008uz}. The result is that
 taking the one-loop corrections to the mass of $\phi_-$ implies that
 the final state $W \tilde l_L$ dominates the decays of $\phi_-$ such
 that all other decay modes have a branching ratio of at most
 $10^{-6}$. This, in turn, implies a total width in the order of meV as
 can be seen in \fig{plot_decay} for the two limiting cases $F\ll M^2$ and $F\simeq M^2$
with dramatic consequences for the
 dark matter considerations as discussed in the next section. As can be seen, due to the
inclusion of $W^-$ final state the width increases such, that the decay temperature
of $\phi_-$ is above the freeze out temperature of the gravitino.

In models with broken $R$-parity one might wonder if the NLSP decays
might yield additional entropy production after the freeze-out of the
gravitino. However, in this model a lower bound on the corresponding
couplings exist due to requirement of correctly explaining neutrino
physics \cite{Porod:2000hv,Hirsch:2002ys,Hirsch:2003fe}. In case of a
neutralino NLSP there are regions in parameter space where it is
long-lived enough to see a displaced vertex at LHC
\cite{deCampos:2007bn} but, nevertheless, it is too short-lived to be of
importance for cosmological considerations. A stau NLSP decays even
faster and, thus, does not play a role either.

\section{Dark matter}
\label{sec:dm}

\begin{figure}[t]
\begin{minipage}{12cm}
\includegraphics[scale=0.5]{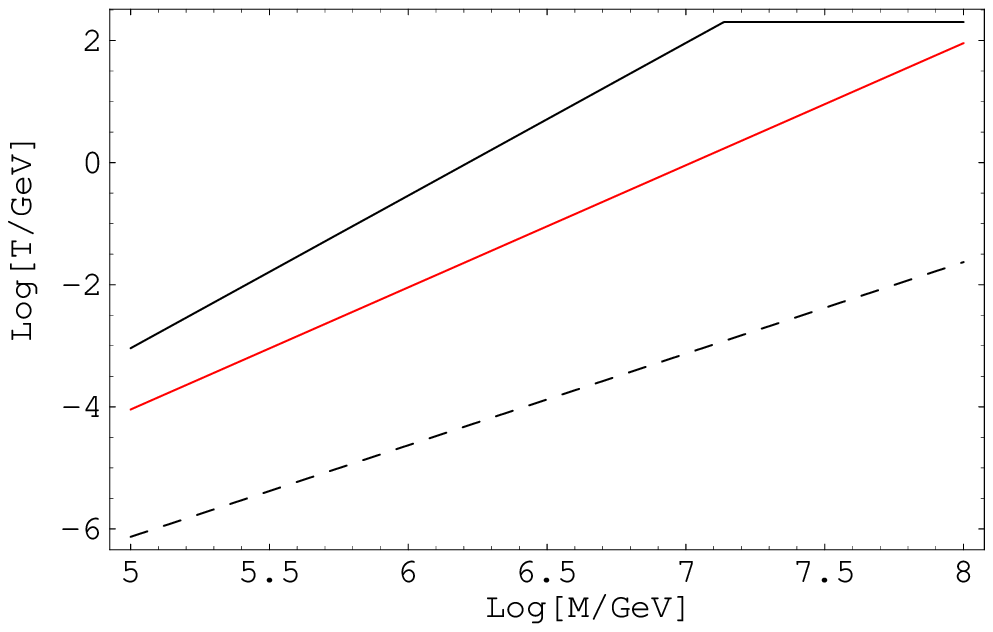} 
\hfill
\includegraphics[scale=0.5]{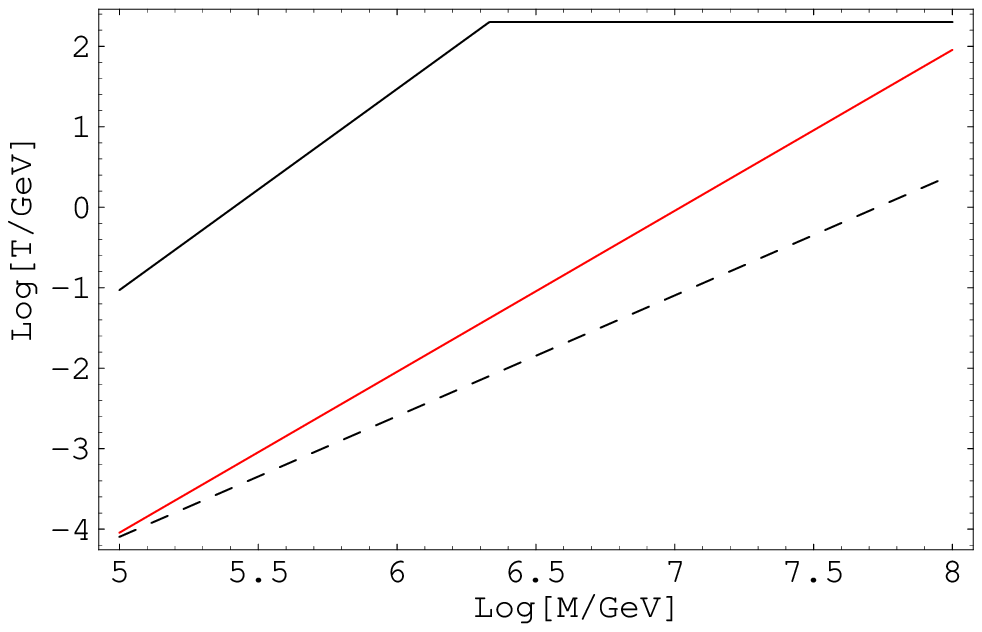} 
\end{minipage}
\caption{Comparison of the the decay temperature of the messenger to the freeze out temperature of the gravitino (red line). The black, solid line shows the temperature including the decay in $W^-$, the dashed line the temperature without this channel. Left plot for $M^2 \gg F$ and right plot for $M^2 \cong F$, both for $k=1,f=1$. } 
\label{fig:plot_decay}
\end{figure}

\begin{figure}[t]
\begin{minipage}{14cm}
\includegraphics[scale=0.6]{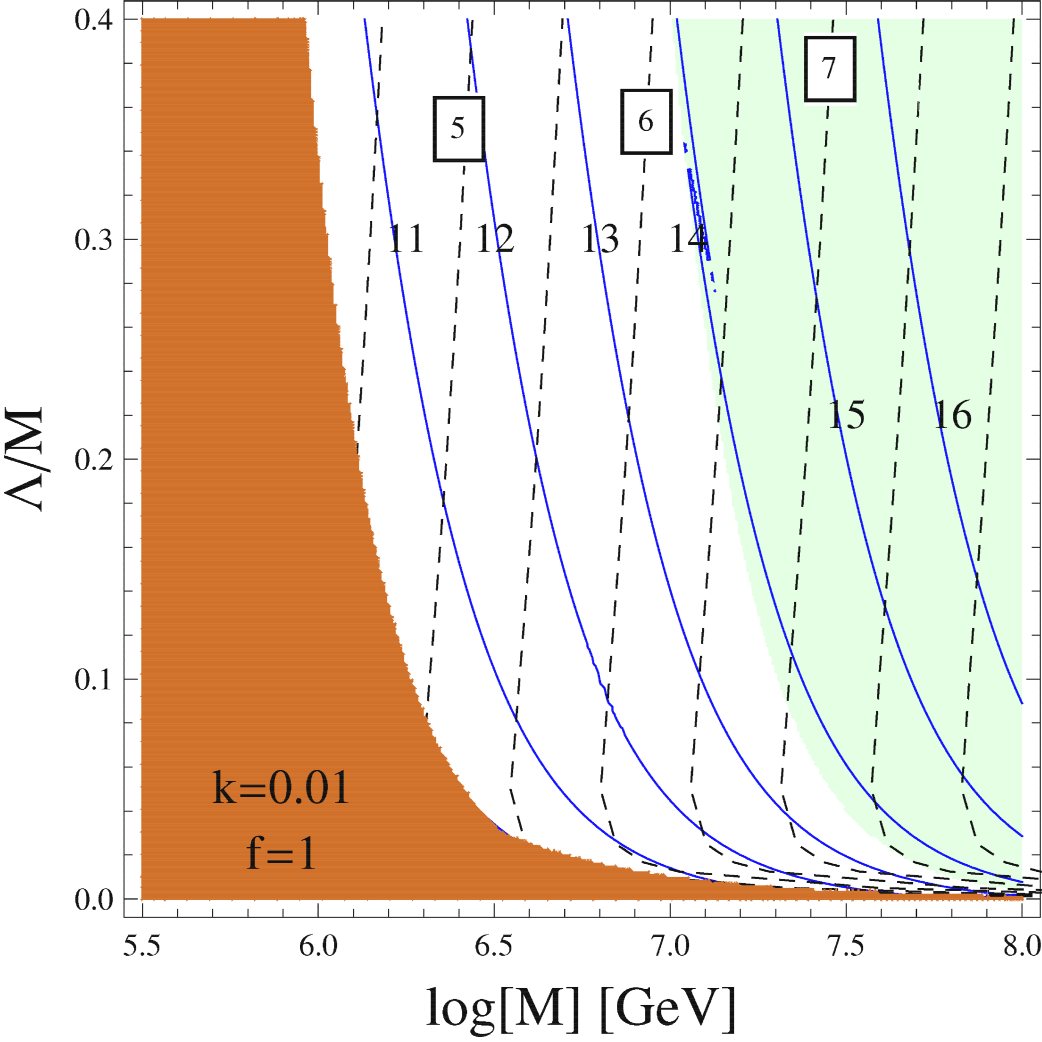} 
\hfill
\includegraphics[scale=0.6]{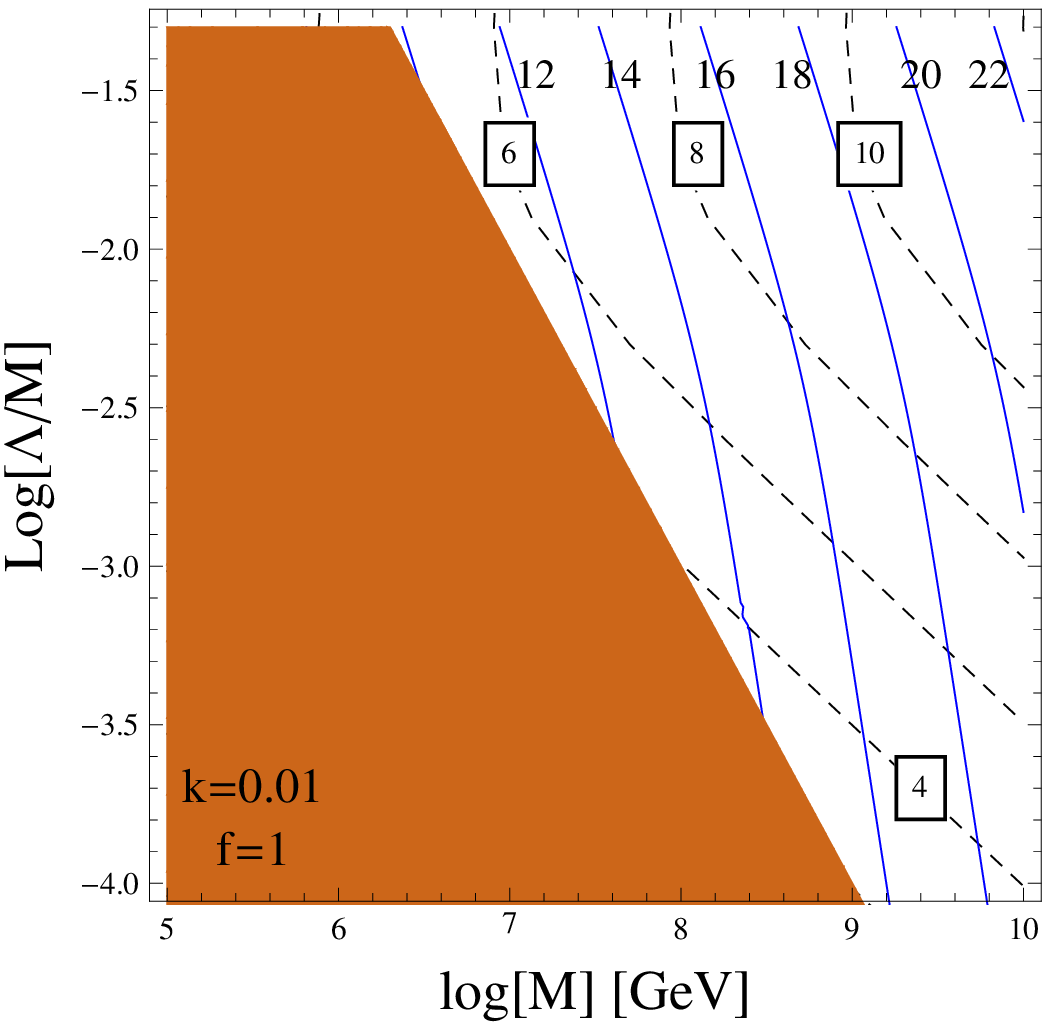}
\end{minipage}
\caption{Comparison between the calculated relic density for the
  lightest strongly interacting messenger scalar (black dashed line)
  and needed relic density for a sufficient dilution of the gravitinos
  (blue solid line). The labels show $\log \Omega$. The right plot is a zoom of the area with
  $\Lambda/M$ below 0.1 of the left plot. The brown area is excluded
  by Lyman-$\alpha$ forest observations and the condition $\Lambda > 10^5 \mbox{GeV}$ because of LEP bounds for SUSY masses. In the green area one gets $m_{3/2} \ge 1$~MeV. }
\label{fig:plot_needed}
\end{figure}

Owing to the mixing term in eq.~(\ref{eq:deltaW}), the lightest messenger
particle ceases to be stable but decays into SM and SUSY particles. The
ratio of the entropy before and after the decay of the messengers is
estimated by \cite{Kolb:1990vq}
\begin{equation}
\frac{s_{\mbox{after}}}{s_{\mbox{before}}} \simeq \frac{4}{3}\frac{Y m}{T_D}
\end{equation}
and leads to a dilution of the relic density of particles already frozen out. 
The question is now whether one can get sufficient dilution after taking
into account the additional decay mode $\phi_-\to W \tilde l_L$.  The
freeze-out temperature of the gravitino is given by
\cite{Giudice:1998bp}
\begin{equation}
T_{3/2} = 0.62 \frac{m^2_{3/2} M_P \sqrt{g_\star}}{\alpha_S m^2_{\tilde{g}}} .
\end{equation}
The gluino mass is given by 
\(m_{\tilde{g}} \simeq \frac{\alpha_s}{4 \pi} \Lambda\)
\cite{Giudice:1998bp} in GMSB.  As we  only need  a rough estimate 
of this temperature, we use eq.~(\ref{eq:m32}) and
set all parameters of \(O(1)\) to 1. As a result, we get \(T_{3/2}\)
as a
function of the messenger mass \(M\): \(T_{3/2} \simeq 10^{-13} \frac{M^2}{k^2 \mbox{GeV}}\).
Comparing this  to the decay temperature of the scalar
messengers in \fig{plot_decay}, one sees that the $SU(2)$ doublet messenger 
always decays before the freeze-out of the gravitinos has taken place.
This effect is independent of the choice of the free parameters and of
the fact that we have considered only one pair of messengers thus far.  As
mentioned above, also the lightest $SU(3)$ messenger is stable in the
limit of conserved messenger number and once this number is broken it
decays according to eq.~(\ref{eq:gamma_neut}) dominantly into a
$d$-quark and a gluino with the correct life time. However, although
the life time is of the required order of magnitude, the number density
is too small to provide sufficient entropy. This can be seen in
\fig{plot_needed}, where the necessary relic density for sufficient
dilution is compared with the obtainable density. For tiny ratios of
\(\frac{\Lambda}{M}\), the annihilation in gravitinos becomes dominant,
which is reflected in the strong bending of the dashed lines in the
plot to the right of \fig{plot_needed},  and inhibits a bigger
dilution. Since the interactions of the Goldstino are proportional to
\(m_{3/2}^{-1}\) the situation gets worse for larger values of \(k\).
The sharp edges of the blue lines in the left plot arise because the degrees of freedom change,  when the decay temperatur crosses a mass threshold, as can be seen from eq.\ (\ref{eq:relic}). 
Note that in \fig{plot_needed} we have covered for completeness
a much larger part of the parameter
space than the one accessible at LHC: mainly the parameter space shown in the right
figure has a sufficiently light spectrum whereas in most  the left plot the spectrum is too
heavy for detection at LHC. Moreover, in the green area one finds a gravitino mass
larger then 1 MeV and, thus, one can avoid the problem of over-closing the universe
if the reheat temperature is sufficiently low \cite{Steffen:2008qp}.

Up to now we have assumed that $f$ is an $O(1)$ parameter. As
mentioned above it results from a non-minimal K\"ahler potential and
naturally is \(O(1)\) \cite{Jedamzik:2005ir}.  As the widths of the
lightest messengers are proportional to $f^2$ one might take smaller
values to increase the life-time of the $SU(2)$ messenger. However,
this does not work in practice, because the life time of the
$SU(3)$ messengers also increase and once $f$ is of $O(10^{-2})$ this
destroys the successful BBN predictions: they decay at
temperature in the low keV range and below. At this time they dominate
the energy density of the universe implying that the energy injection
due to their decays would dissociate BBN products to a very high
extent.  In addition, the reheating temperature of this decay is
related to the decay temperature and the dilution factor by
\(T^{SU(3)}_{RH} = T^{SU(3)}_D \sqrt[3]{\Delta^{SU(3)}}\).  For the
calculation of the dilution factor we used MicrOmegas \cite{Belanger:2006is}.  This reheating
is too low to restart BBN again, especially since the number density
of the \(SU(3)\) messengers gets also diluted by the decay of the
\(SU(2)\) messengers. To obtain a rough impression of the reheating
temperature, we can approximate \(Y_{SU(3)} \simeq Y_{SU(2)}\) and
\(M_{SU(3)} \simeq M_{SU(2)}\) leading to
\begin{equation}
T^{SU(3)}_{RH} = T^{SU(3)}_{RH} \left(\frac{Y_{SU(3)}}{Y_{SU(2)}} \frac{M_{SU(3)}}{M_{SU(2)}} \frac{T_D^{SU(2)}}{T_D^{SU(3)}}\right)^{\frac{1}{3}} \simeq 
T^{SU(3)}_D \left(\frac{g}{g_3} \frac{M}{M_W}\right)^{\frac{1}{3}}, 
\end{equation}
i.e. the reheating temperature remains in the keV range.

Therefore, it is also not possible to adjust the parameters in a way
that a combined dilution of both decays leads to a correct Gravitino
abundance and circumvents BBN bounds, because the ratio of the yields
and of the decay temperatures is fixed by the gauge couplings, the
messenger mass $M$ and the $W$ mass.

\section{Variations of the messenger sector}

So far, we have only considered a minimal GMSB scenario with one
messenger pair transforming as \(5\) and \(\bar{5}\) under
\(SU(5)\). We have seen, that the \(SU(2)\) messengers could in
principle produce the required additional entropy if we fine-tune
their decay temperature to the minimal value allowed by BBN by
reducing the parameter $f$.  But the \(SU(3)\) messengers decay in
this scenario much later, and this scenario is  therefore ruled out by
BBN. This result is independent of the concrete mixing term, e.g., it does
not change if instead one takes higher-dimensional operators inducing
the mixing of the messenger particles with the MSSM particles. We have
checked that for all cases proposed in \cite{Jedamzik:2005ir}, which
contains an exhaustive list of possibilities, the same conclusions
hold. The picture changes a little bit if the messenger multiplet
contains $U(1)$ charged particles which are \(SU(3)\times SU(2)\)
singlets, i.e.\ \(\tilde{e}_R\)-like messengers. They have a relic
density comparable to \(\tilde{v}_L\)-like messengers, but their decay
width is not enhanced by decays into vector-bosons.  Therefore, these
messengers can have the smallest decay width of all
messengers. However, in out numerical studies we have only found tiny
isolated regions in parameter space rendering this possibility highly
unattractive as this requires quite some fine-tuning of the
parameters.

\begin{figure}[t]
\begin{minipage}{14cm}
\includegraphics[scale=0.6]{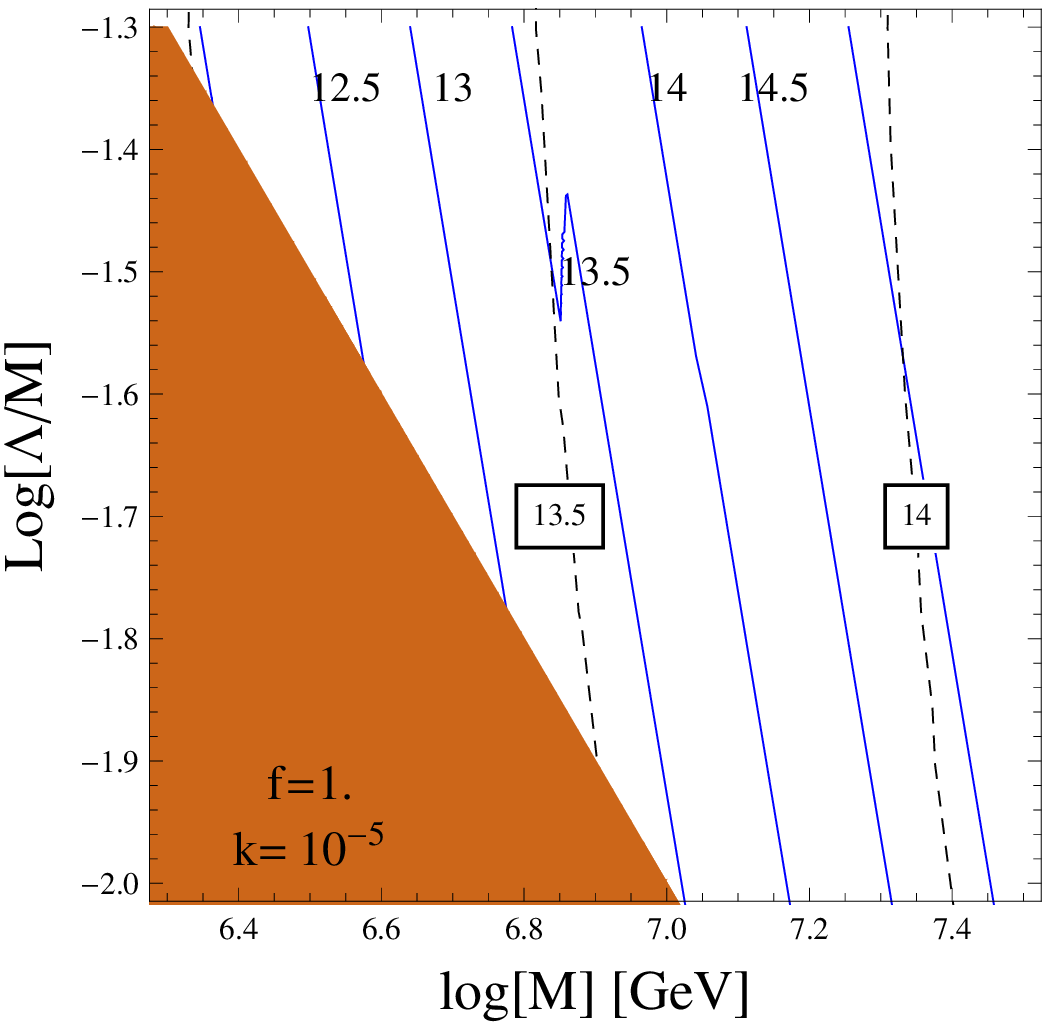} 
\hfill
\includegraphics[scale=0.6]{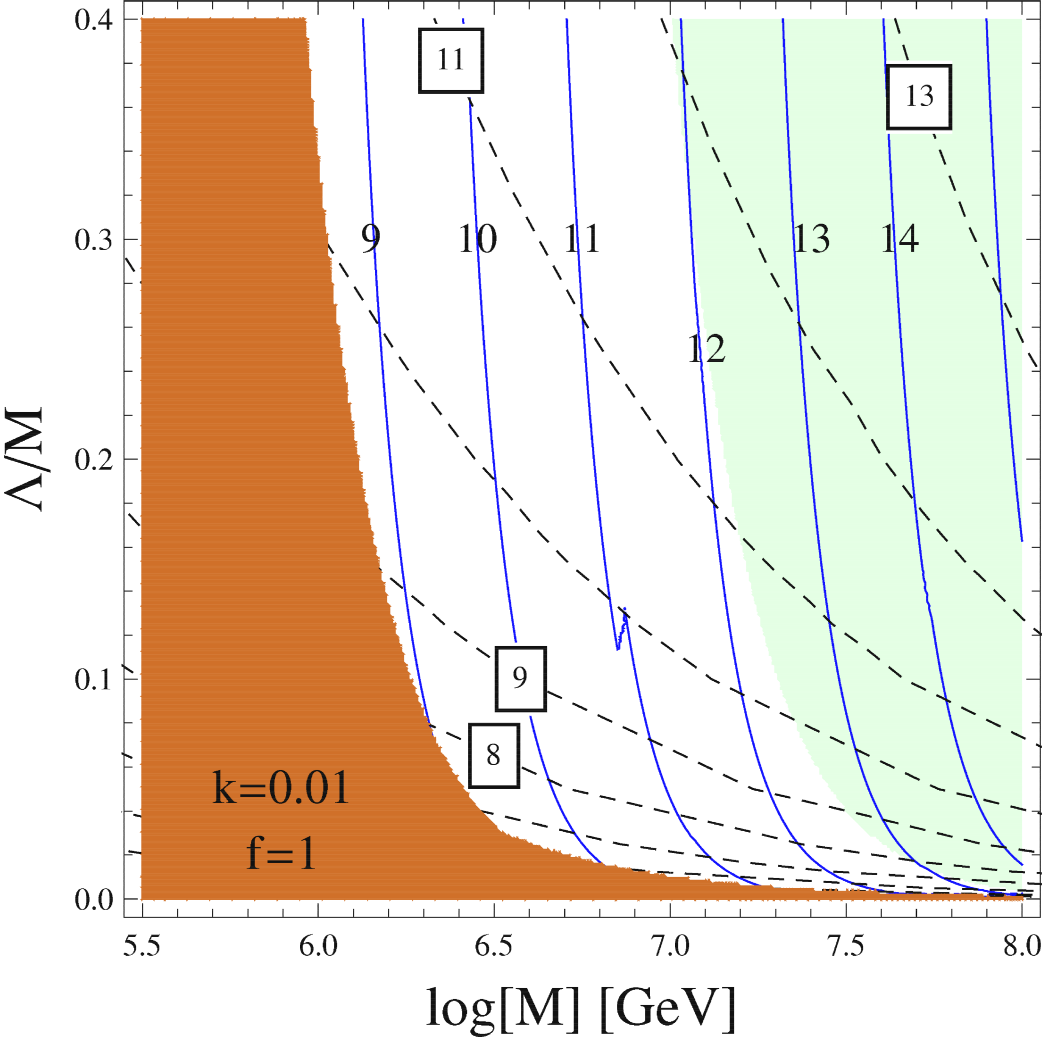} 
\end{minipage}
\caption{Comparison between the calculated relic density for gauge
  singlet Messenger (black dashed line) and needed relic denisty for
  sufficient delution (blue solid line). The labels show $\log \Omega$.  Left plot for a ratio $r$
  smaller 0.01 and $k=10^5$ , right plot for larger $r$ and $k$.  The
  brown area is excluded by Lyman-$\alpha$-forest observation and the condition 
$\Lambda >10^5 \mbox{GeV}$. In the green area one gets $m_{3/2} \ge 1$~MeV.}
\label{fig:vR}
\end{figure}

The last possibility is to consider a messenger sector containing
singlets under all gauge groups (\(\tilde{v}_R\)-like) as it appears
for example in the 16-plet of $SO(10)$.  This was already analyzed in
\cite{Lemoine:2005hu} where it has been shown that this scenario works
in principle.  Their analysis was based on annihilation due to loop
vertices, which are dominant if the spurion mass is of the order of
the messenger mass and \(M\simeq 10^6\). We reconsidered this
scenario for the case that all particles of the hidden sector
are heavier than the messengers. As a consequence, the dominant
interaction of the messengers is always due to the Goldstino component
of the gravitino.  Their relic density and their decay width depend
only on this interaction and, thus, the decay in MSSM particles is
given by
\begin{equation}
\Gamma = \frac{1}{16\pi}\left(\frac{m_-^2}{m_{3/2} m_{Pl}}\right)^2 \delta^2 m_- .
\end{equation}
The result of our numerical studies is that for the case \(M^2 \gg
F\) and \(f=1\) this scenario works for very small values of \(k\)
(\(O(10^{-5})\)) and messenger masses of \(O(10^7)\), as can be seen in
\fig{vR} on the left. For larger values of \(k\) or \(M\), the
annihilation is too effective and the additional entropy production
too small. This can be partly compensated by reducing \(f\) as the
dilution behaves like \(\Delta \sim k^{-4} f^{-1}\) and we reach the
BBN bound very fast. However, if we assume \(\Lambda \simeq M\), one
finds solutions for larger \(k\) as can be seen figure \ref{fig:vR}
right. The reason is that the annihilation in Goldstinos, a
t-channel interaction, is suppressed by the large mass splitting. The
messengers are in the PeV-range and the gravitino mass is about 10 keV,
i.e.\ it is still warm dark matter but not in conflict with the
Lyman-\(\alpha\) observations. Note that in contrast to charged
messenger particles a $\tilde{v}_R$ messenger does not receive large
one-loop corrections to its mass due to gauge interactions.
Note that as in \fig{plot_needed} we have covered in \fig{vR} for completeness
a much larger part of the parameter
space than accessible at LHC.

\section{Conclusions}
\label{sec:conc}

We have reconsidered the question of dark matter in gauge mediated
SUSY breaking.  Assuming that the reheat temperature is large enough
so that the gravitino gets into thermal equilibrium, we 
find that in the simplest models it is not possible
to obtain the correct amount of dark matter once all contraints are
taken into account: observation of the Lyman-\(\alpha\) forest implies
that a gravitino with a mass above about 8 $keV$ is needed which
yields too much dark matter if only the standard history of the
universe is considered. The mechanisms to produce
additional entropy via messenger number violating terms proposed so
far do not work in
scenarios where the messenger come in complete $SU(5)$ representations
as a consequence of two facts which have been overlooked so far: (i)
the decay of the messenger particle has to occur after electroweak
phase transistion and thus decays into $W$-bosons have to be
considered and (ii) for each represenation of the SM gauge group one
has a lightest messenger scalar which would be stable if messenger
number were a conserved quantum number. Roughly one finds two main
possiblities: the $SU(2)$ messengers decay too fast and the $SU(3)$
messengers do not produdce sufficient entropy to dilute the gravitinos,
or the $SU(2)$ messengers decay sufficiently late and the $SU(3)$
messengers destroy the predictions for BBN. However, by sufficiently lowering
the reheat temperature one can avoid this problem once the gravitino mass is
in the range of 1 MeV and above.

Although we have mainly focused on terms of the form $f m_{3/2}
\hat{\Phi} \hat{\bar{5}}_M$ in detail, one can easily extend this
discussion to other scenarios. We have checked that for all cases
proposed in ref.~\cite{Jedamzik:2005ir} which contains an exhaustive
list of possibilities, the same conclustions hold but for a tiny region
in parameter space where a $\tilde e_R$ like messenger is the lightest
messenger particle. Moreover, we have also seen that scenarios with a
SM gauge singlet lightest messenger, occuring for example in
16-plet of $SO(10)$ models, are still viable. \\
In this study, we have assumed that the gravitino constitutes all of
the dark
matter. If one relaxes this assumption, one has to invoke scenarios
where the gravitino has a mass below 60 eV (see (\ref{eq:relic})) if at most 60\% of the dark
matter is allowed to be warm dark matter \cite{Boyarsky:2008xj} where
this limit has been found in case of a 5 KeV gravitino. Of course the
mass gets further constrained if the allowed warm dark matter
contribution has to be smaller. Taking into account the existing
collider constraints on chargino, slepton and Higgs masses
\cite{Amsler:2008zzb}, we get a lower bound of about 3 eV on the
gravitino mass. In the mass range between 3 and 60 eV, the NLSP life
time is such that in principle a displaced vertex can be measured at the
LHC  which scales roughly like $m_{3/2}^2$. Thus, one can get
information on the gravitino mass once the properties of the decaying
particle are known. This in turn would be a clear indication that the
gravitino cannot form all of the observed dark matter. In case of a
gravitino with mass of $O(1)$~eV, one might even find imprints in the
reconstructed CMB lensing potential \cite{Ichikawa:2009ir}.

\section*{Acknowledgments}

This work is supported by the DFG Graduiertenkolleg GRK-1147 and by
 the DAAD, project number D/07/13468.

\input{lit_paper.tex}

\end{document}

%% file: lit_paper.tex
\bibliographystyle{h-physrev}